\documentclass[10pt]{article} % For LaTeX2e
\usepackage[accepted]{rlc}
\usepackage{amssymb}            % Defines common symbols like \mathbb R
\usepackage{mathtools}          % Extends amsmath, providing common math tools
\usepackage{mathrsfs}           % Enables \mathscr, which can work in cases that \mathcal does not
\usepackage{graphicx}           % For including images
\usepackage{subcaption}         % Allows for the use of subfigures and subcaptions
\usepackage[space]{grffile}     % For spaces in image names
\usepackage{url}                % For displaying URLs

\usepackage{float}
\usepackage{algorithm}
\usepackage{algpseudocode}
\graphicspath{ {./figures/} }

\title{Hierarchical Multi-Armed Bandits for the Concurrent Intelligent Tutoring of Concepts and Problems of Varying Difficulty Levels}

\author{Blake Castleman  \\
    bc3029@columbia.edu \\
    Columbia University\\
    Aiphabet, Inc.
    \And
    Uzay Macar \\
    uzay@aiphabet.org \\
    Aiphabet, Inc.
    \And
    Ansaf Salleb-Aouissi \\
    ansafsalleb@columbia.edu\\
    Columbia University\\
    Aiphabet, Inc.}

\begin{document}

\maketitle

\begin{abstract}
Remote education has proliferated in the twenty-first century, yielding rise to intelligent tutoring systems. In particular, research has found multi-armed bandit (MAB) intelligent tutors to have notable abilities in traversing the exploration-exploitation trade-off landscape for student problem recommendations. Prior literature, however, contains a significant lack of open-sourced MAB intelligent tutors, which impedes potential applications of these educational MAB recommendation systems. In this paper, we combine recent literature on MAB intelligent tutoring techniques into an open-sourced and simply deployable hierarchical MAB algorithm, capable of progressing students concurrently through concepts and problems, determining ideal recommended problem difficulties, and assessing latent memory decay. We evaluate our algorithm using simulated groups of 500 students, utilizing Bayesian Knowledge Tracing to estimate students' content mastery. Results suggest that our algorithm, when turned difficulty-agnostic, significantly boosts student success, and that the further addition of problem-difficulty adaptation notably improves this metric.
\end{abstract}

\section{Introduction}
With a rise in technologically-assisted education in the twenty-first century, intelligent tutoring systems have received significant popularity across various domains (\citeauthor{ITSReview2}, \citeyear{ITSReview2}; \citeauthor{ITSReview1}, \citeyear{ITSReview1}). Though intelligent tutoring systems spread across multiple domains of artificial intelligence, recent literature has argued that reinforcement learning models are optimal for adaptive learning environments (\citeauthor{yanRLBest}, \citeyear{yanRLBest}). This is due to the ability of reinforcement learning to best sequence actions in uncertain and dynamic landscapes without prior data on students or educational content (\citeauthor{kaelblingRL}, \citeyear{kaelblingRL}). As a result, abundant resources have been allocated toward the research and expansion of reinforcement learning applications in education.

Within the realm of reinforcement learning, the multi-armed bandit (MAB) stands out as a distinctive framework for addressing and deploying exploration-exploitation trade-offs in a generalized manner. In the MAB framework, an agent is equipped with multiple arms, each representing a distinct action it can perform. The objective is to strategically select optimal arms over successive interactions with the environment, given the rewards gained from its prior actions chosen. This strategy aims to maximize cumulative rewards in a transient system.

Intelligent tutoring systems utilizing the MAB framework have since been applied for education (\citeauthor{clementeZPDES}, \citeyear{clementeZPDES}) with various modifications made for performance improvement (\citeauthor{muMABtheory}, \citeyear{muMABtheory}; \citeauthor{segalDifficultyMAB}, \citeyear{segalDifficultyMAB}). However, there exists a significant lack of open-sourced MAB intelligent tutors, which is vital for real-life applications of MAB on pedagogy (observational based on present-day literature), and literature denoting real trial applications of MAB intelligent tutoring systems in real-world learning environments (\citeauthor{muiLackOfMABStudies}, \citeyear{muiLackOfMABStudies}).

In this paper, we seek to alleviate the former dilemma by contributing a deployable, open-source, and state-of-the-art MAB intelligent tutoring algorithm for remote education. First, we describe the learning platform of our nonprofit organization, Aiphabet\footnote{https://aiphabet.org}, designed to informally teach middle and high school students about artificial intelligence. This is the structural framework in which our model will be devised about (Section \ref{sec:platform}). Then, we detail our own MAB intelligent tutor architecture, which combines successful aspects of prior research into MAB intelligent tutoring literature for our organization's platform. Our algorithm utilizes novel \textit{hierarchical MABs}, where separate MAB agents are called to select a concept and dependently select a problem. This algorithm also takes into account predefined problem difficulties from educational supervisors (Section \ref{sec:algorithm}), adapted from prior literature. Lastly, we perform Bayesian Knowledge Tracing (BKT) simulations (Section \ref{sec:simulations}) to provide a theoretical basis for variability between a static expert-defined progression sequence and our hierarchical MAB-defined progression sequence. 

\section{Related Work}
\subsection{Multi-Armed Bandit Intelligent Tutoring Frameworks - Without Difficulty Levels}
The domain of MABs for intelligent tutoring is well defined in the literature (\citeauthor{muiLackOfMABStudies}, \citeyear{muiLackOfMABStudies}) and indeed has been distinguished in previous research. \cite{clementeZPDES} proposed using MABs to recommend activities that optimize the exploration-exploitation trade-off for personalized student progression sequences, developing their algorithm, Zone of Proximal Development and Empirical Success (ZPDES), to deliver problems within students' Zone of Proximal Development (ZPD) (\citeauthor{chaiklinZPDtheory}, \citeyear{chaiklinZPDtheory}). Proceeding literature modified ZPDES for more defined environments, with \cite{muProgram2Tutor} modifying ZPDES to use probabilistic entropy for finding the initial ZPD and \cite{muMABtheory} combining it with the Multiscale Context Model (MCM; \citeauthor{pashlerMCM}, \citeyear{pashlerMCM}) concept forgetting mechanism. 

Another family of MAB intelligent tutors follows \cite{lanCLUB}, who developed upper confidence bound (UCB)-based algorithms (\citeauthor{auerUCB}, \citeyear{auerUCB}) to maintain expected arm rewards with confidence intervals for personalized learning actions (PLAs) that maximize learning. \cite{manickamAdaptedCLUB} built on this with an estimate framework for students' prior knowledge with sparse factor analyses on their previous responses. Additionally, they investigated new policies for selecting PLAs that were adapted for binary-value student correctness rewards.

\subsection{Multi-Armed Bandit Intelligent Tutoring Frameworks - With Difficulty Levels}
Difficulty level adoption has been well observed in modern MAB algorithms with disparate methods utilized. RiARiT, from \cite{clementeZPDES}, utilizes multiple nodes per individual activity, each node representing a more difficulty level than the last, in its progression graph. This allows the ZPD to automatically choose whether it should pursue a proceeding activity or a previously tested activity of higher difficulty. \cite{andersenKnowledgeMatrix} used a novel probabilistic knowledge matrix to facilitate their progression, with rows and columns representing proceeding concepts and difficulty levels respectively. \cite{segalDifficultyMAB} performed difficulty adjustment by linearly scaling the exploration factor and by directly adjusting harder questions' weights based on a student's answer's correctness.

\section{Methodology: Platform Architecture} \label{sec:platform}
Our platform, Aiphabet, is a secondary school informal learning organization for teaching artificial intelligence. The curriculum (\citeauthor{aiphabet}, \citeyear{aiphabet}) is cocurated by many Columbia University faculty members and students who are specialized in both computer science education and learning sciences fields.

\subsection{Platform Section, Concept, and Problem Definition}

We begin by defining educational content sections for students to progress through. In our educational landscape, we define each \textit{section} as consisting of an animated video lecture (approximately three to five minutes long) or a slideshow of course content. Then, following each section is a quizzing stage where students are asked sequential, material-related questions. Finally, after the question sequencing is completed, the student may proceed to another section of their choice, provided they have finished all required prerequisite sections. 

The educational content of a section is comprised of multiple \textit{concepts}. These concepts form a concept progression tree, where an arbitrary concept can become teachable after its prerequisites have been understood and mastered. For example, if a section is titled ``The Perceptron'', corresponding concepts may be ``Biological Inspiration'', ``Classification Function'', and ``The XOR Problem.''

Under each concept, we define \textit{problems} as available questions for examining students' knowledge mastery of said concept. Difficulty levels are scores to rate problem complexity by a domain expert within the range $d \in [1,5]$ where higher levels denote more difficult questions. These parameters, however, are not given to the student but are used for internal algorithm calculations (Section \ref{difficulty-usage}). \label{difficulty_score}

\section{Methodology: ZPDES Foundation for Algorithmic Progression} \label{sec:algorithm}

A MAB framework is used to select the concepts for students to be quizzed on in order to best solidify the educational content of a section. However, as multiple questions with varying difficulties exist for quizzing a particular concept with, it is not sufficient to have a single MAB instance selecting both the concepts and problems at hand.

We solve this problem by implementing a progression algorithm embedded with multiple MAB agents to explore the exploration-exploitation trade-off for students among both concepts and problems. First, a single, MAB for concept selection (high-level decision; known as the concept MAB) chooses a concept within the given section to quiz the student on, and then a corresponding MAB instance for the selected concept (low-level decision; known as the problem MAB) chooses a problem from that concept's question bank to give to the student (see Appendix \ref{sec:mab_schematic} for a visual representation). Initializations for the proceeding parameters can be found in Appendix \ref{zpdes-maple-hypertune}. We open-source our code for educators and researchers to implement and build upon.\footnote{https://github.com/b-castleman/hierarchical-mab-tutoring}

\subsection{ZPDES Multi-Armed Bandit Design Foundation}
The base algorithm used for our MAB adaptation extends the \cite{muMABtheory} MAB intelligent tutoring implementation. \cite{muMABtheory}'s algorithm combines the ZPDES algorithm and MCM model for characterizing concept forgetting in a single study session. 

The ZPDES algorithm aims to select student activities within a ZPD frontier. The ZPD is an educational psychology idea that hypothesizes that optimal student activities should be difficult enough for a student to be challenged by but not outside of a student's current problem-solving abilities (\citeauthor{chaiklinZPDtheory}, \citeyear{chaiklinZPDtheory}). As a result, concepts and problems within a student's ZPD can challenge students while preventing frustration, which increases motivation and student engagement.

ZPDES is a MAB algorithm designed to deploy optimized teaching sequences by exploring the exploitation-exploration trade-off for student activities. By keeping track of belief states (denoted as unmastered or mastered) for each activity, it proposes unmastered activities for students to solve that are within their ZPD (i.e., unmastered activities that have mastered prerequisites).  This occurs by first computing the learning progress $r_{a,t}$, also known as the reward, for a certain activity $a$:
\begin{equation}r_{a,t} = \sum_{k=t-L/2}^t \frac{C_{a,k}}{L/2} - \sum_{k=t-L}^{t-L/2} \frac{C_{a,k}}{L-L/2}\end{equation}
where $C_{a,k}$ is the correctness of the exercise given at time $k$, $L$ is the history length, and $t$ is the current time being analyzed for mastery (\citeauthor{clementeZPDES}, \citeyear{clementeZPDES}; \citeauthor{muMABtheory}, \citeyear{muMABtheory}). The equation compares the success of the last $L/2$ samples with the preceding $L/2$ samples to approximate a performance gradient for this activity. In our implementation, if the designated history length is greater than the current activity history, we assume the correctnesses of any hypothetical earlier exercise to be zero.

The next problem is then selected by converting these rewards into weights ($w_a$) for probabilistic selection. In prior literature (i.e., \citeauthor{clementeZPDES}, \citeyear{clementeZPDES}), MAB rewards are often used to update weights with $w_a := \beta w_a + \eta r_{a,i}$, where $\beta$ and $\eta$ are hyperparameters for rate updates. However, we instead choose to follow \cite{muMABtheory}'s weight update formulation, an averaging of all previous rewards, in order to decrease the hyperparameters required for system tuning:
\begin{equation}
w_a = \frac{1}{n_a} \sum_{k=1}^{n_a} r_{a,k} \label{eq:weight_creation}
\end{equation}
where $n_a$ is the total number of times activity $a$ has been presented to the student (\citeauthor{muMABtheory}, \citeyear{muMABtheory}). Finally, the activity weights are normalized based on the current activities in the ZPD and then are introduced a factor of exploration probabilities for probabilistic selection:
\begin{equation}
w_{a,n} = \frac{w_a}{\sum_{a \in ZPD} w_a},\quad p_a = w_{a,n}(1-\gamma) + \frac{\gamma}{|ZPD|} \label{eq:exploration}
\end{equation}
where $\gamma$ is the exploration rate hyperparameter and $|ZPD|$ is the current size of the ZPD (\citeauthor{clementeZPDES}, \citeyear{clementeZPDES}; \citeauthor{muMABtheory}, \citeyear{muMABtheory}). The activity is then randomly picked from the ZPD based on each probability $p_a$.

Once the average correctness over the last $L$ attempts surpasses a threshold hyperparameter $h$ ($h > \frac{1}{L} \sum_{k=n_a-L+1}^{n_a} C_{a,k}$), we choose to change the belief state of the activity to mastered and remove it from the ZPD. This is consistent with \cite{muMABtheory}'s approach but instead accounts for the correctness variable rather than the net accuracy, which has particular implications for the concept MAB implementation that will be discussed in Section \ref{difficulty-usage}.

\subsection{MCM Algorithm Adaptation}

In \cite{muMABtheory}, the MCM model is integrated into ZPDES to approximate students' decaying memory traces over time (\citeauthor{pashlerMCM}, \citeyear{pashlerMCM}). We choose to incorporate this as well for our algorithm to parallel our work with present literature.

A memory trace $x_{a,i}$ (induced memory change) of an activity $a$ the $i$th time after it has been seen by a student decays with the time after its activation. This can be modeled according to the equation:
\begin{equation}
x_i(t + \Delta t) = x_i(t) e^{ (\frac{-\Delta t}{\tau_i}) }
\end{equation}
where $\tau_i$ is the decay time constant with the constraint $\tau_i < \tau_{i+1}$, $t$ is the activation time, and $\Delta t$ is the time since activation (\citeauthor{muMABtheory}, \citeyear{muMABtheory}; \citeauthor{pashlerMCM}, \citeyear{pashlerMCM}). The probability of receiving an activity is related to the memory strength $s_{a,t}$ of a problem $a$ after it has been seen $n$ times:
\begin{equation}
s_{a,t} = \frac{1}{\Gamma_n} \sum_{i=1}^{n} \xi_i x_{a,i}(t) \text{,\quad where} \quad \Gamma_n = \sum_{i=1}^{n} \xi_i
\end{equation}
where each $\xi_i$ is a weight for each memory trace $x_{a,i}$ (\citeauthor{muMABtheory}, \citeyear{muMABtheory}; \citeauthor{pashlerMCM}, \citeyear{pashlerMCM}). 

Lastly, to integrate MCM with ZPDES, the set of activities available for selection is extended to include a set $M$, the activities believed to have been previously learned. First, all activities in the ZPD have weight calculations and updates as previously described. However, before probabilistic selection, activities in the learned set obtain weights through the equation:
\begin{equation}
w_a = m_m \text{max}(0,m_t-s_a), a \in M
\end{equation}
where $m_t$ is a memory threshold and $m_m$ is a memory multiplier for describing the effects of forgetting (\citeauthor{muMABtheory}, \citeyear{muMABtheory}). Then, normalization again occurs for the set of all possible problems before probabilistic selection.

\begin{figure*}
\begin{center}
\includegraphics[width=0.95\linewidth]{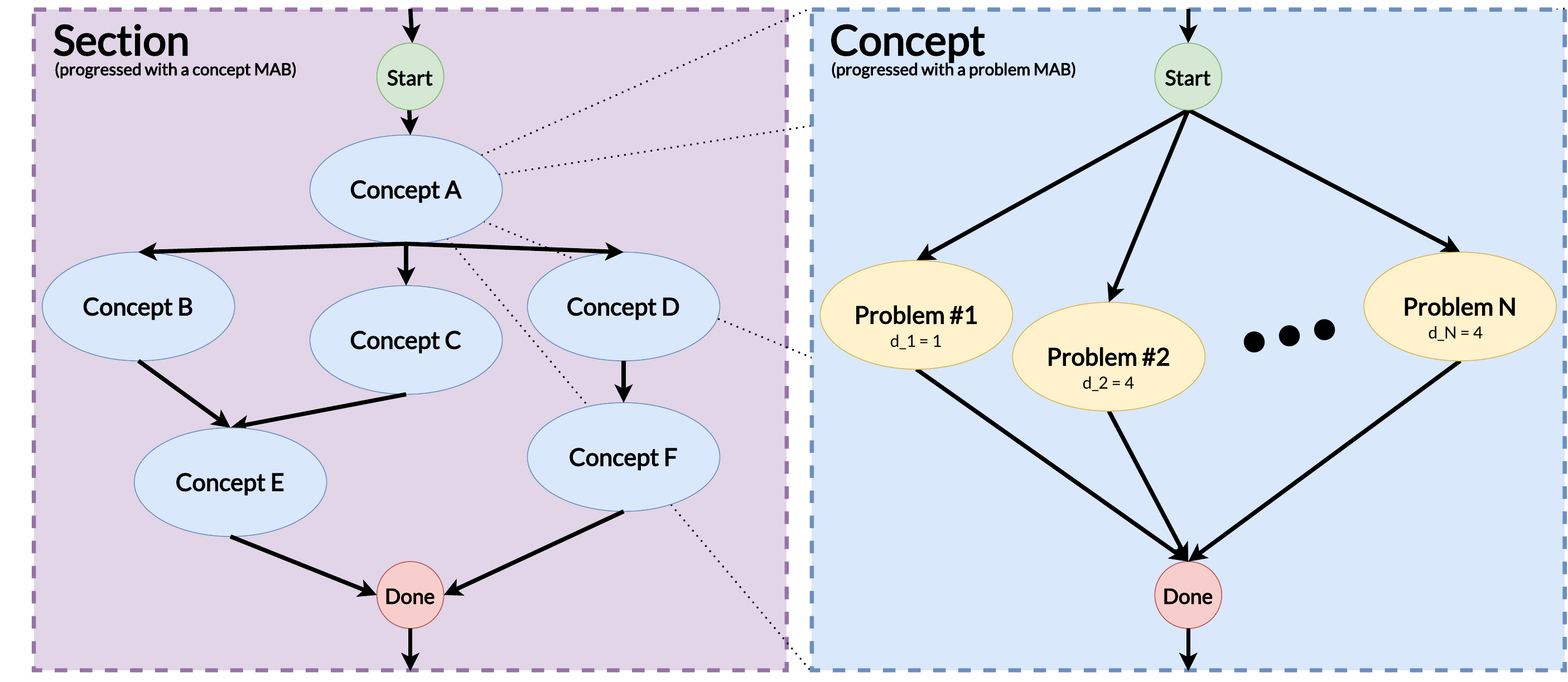}
\caption{\label{fig:progression-trees}Progression tree examples for both the conceptual MAB and the problem MAB. Note that there are separate problem progression trees for each concept in a given section.}
\end{center}
\end{figure*}

\section{Methodology: High-Level Concept Multi-Armed Bandit}

With a MAB design foundation formed, we will now proceed to describe our induced MAB hierarchy and difficulty extensions to the system, beginning with the MAB for concept selection. 

\subsection{Concept Progression Trees}

Each section on the platform has multiple concepts that are taught by the educational content. To solidify student knowledge, the MAB for concept progression is designated to quiz students on the entire section that they have learned, beginning with the root prerequisite concepts and leading to the more advanced concepts previously taught. For a concept to be taught, all prerequisites that lead to it must have a mastered belief state. Figure \ref{fig:progression-trees} details an example of how a concept progression tree may be defined, noting that concepts can run in series, parallel, and share postrequisites.

\subsection{Problem Difficulty}

As the difficulty of the problems solved (Section \ref{difficulty_score}) affects the estimated knowledge trace for each concept, we choose to incorporate the difficulty of completed problems into the reward calculation. 

To account for difficulty, when a problem is completed, we update response correctness according to: 
\begin{equation}
    C_{a,k} = C_{a,k} * (\sigma(d-3) + \frac{1}{2})
\end{equation} \label{difficulty-usage}
 where $d$ is the difficulty of the problem and $\sigma(x)$ is the sigmoid function. We choose to perform $d-3$ to center the difficulty scale along the sigmoid function (as $d \in [1,5]$). The sigmoid function then scales the value to be within $[0,1]$, which is ideal as it contains asymptotic behavior but exhibits sigmoidal growth. Finally, lifting the value by $\frac{1}{2}$ results in the multiplier to be within $[0.5,1.5]$. This multiplier generation therefore allows the ZPDES algorithm to reward more complex question answers generously while maintaining frugality for simple problems.

\section{Methodology: Low-Level Problem Multi-Armed Bandit}
After concept selection, a problem MAB unique to the selected concept chooses the problem for student presentation. This allows for the concept's MAB to obtain parameters specific to the concept and resultant question bank. We will now proceed to describe our difficulty extensions to the system, beginning with the MAB for concept selection. 

\subsection{Problem Progression Trees}

Each concept has multiple associated problems from that concept's question bank that make up a problem progression tree. These trees have a depth of one and consist only of edges in parallel from a root to each associated problem from that concept's question bank. As a result, the MAB framework is capable of intelligently selecting all problems with trade-off considerations when the concept is selected. Figure \ref{fig:progression-trees} demonstrates this problem progression tree design on an abstract concept.

\subsection{Initial Problem Difficulty Integration}

As all problems have an initial equal probability for selection through the base design of our MABs framework, we seek to integrate the difficulty of the problems in order to skew the initial weights. Therefore, very easy or hard problems can be initially discouraged before information on the student in the current section is available.

To accomplish this, we introduce a problem multiplier variable $m_{a}$ that is multiplied onto the problem weight $w_a$ after the weight calculation (Equation \ref{eq:weight_creation}) and before the normalization and exploration steps (Equation \ref{eq:exploration}). This allows us to skew resultant weights according to difficulty adjustments. 

We choose to initialize the multipliers $m_{a}$ for each problem within a concept according to: \begin{equation}
    m_a = e^{-\frac{(d - 3)^2}{\xi}}
    \label{eq:initialProblemDifficultyIntegration}
\end{equation}
where $d$ the difficulty of the problem (as aforementioned) and $\xi$ is a hyperparameter for initial problem weight skewing. After performing a centering along $d=3$, this equation, a Gaussian function, allows intermediate problems to be encouraged while discouraging problems of increasing simplicity or difficulty. Thereby, the MABs framework is more likely to initially suggest a problem meeting these conditions.

\subsection{Transient Problem Ranking Integration}

As student correctness data is collected on problems within the given concept, students will demonstrate skill aptitudes which, even if the belief state of the given concept is unmastered, must be perceived and processed for further recommending difficulty rankings. Thereby, students that have completed difficult problems can be recommended problems of greater difficulty and students unable to complete difficult problems can be recommended problems of lesser difficulty.

% To accomplish this, we incorporate parts of the Multi-Armed Bandits based Personalization for Learning Environments (MAPLE) algorithm \cite{segalDifficultyMAB}, which combines difficulty ranking with MABs for this exact problem. MAPLE updates problem weights according to the following method:
% \begin{algorithm}[H]
%     \caption{MAPLE Weight and Exploration Update}
%     \begin{algorithmic}
%     \State \textbf{Data:} Passing grade threshold $\eta$, exploration rate $\gamma$, normalization factors $\alpha_1,...,\alpha_4$, question $q_a$
    
%     \State Get student grade $g_s$ after solving question $q_a$
%     \If{$g_s > \eta$}  
%         \State Increase weights for questions more difficult than $q_a$:
%         \State $w_a = \alpha_1 e^R w_a$
%         \State Increase exploration factor: $\gamma = \alpha_2 \gamma$
%     \Else
%         \State Decrease weights for questions more difficult than $q_a$:
%         \State $w_a = \alpha_3 e^R w_a$
%         \State Decrease exploration factor: $\gamma = \alpha_4 \gamma$
%     \EndIf
%     \end{algorithmic}
    
% \end{algorithm}

To accomplish this, we incorporate and modify the Update Question Grade calculation step from the Multi-Armed Bandits based Personalization for Learning Environments (MAPLE) algorithm (\citeauthor{segalDifficultyMAB}, \citeyear{segalDifficultyMAB}), which combines difficulty ranking with MABs for this exact problem. MAPLE updates problem weights according to the following method. For our implementation, we choose to make various modifications to this step in MAPLE for our application, which can be found in Appendix \ref{sec:maple-changes-made}. Along with tuning of the initial parameters (Appendix \ref{zpdes-maple-hypertune}), the final implementation of the problem ranking integration therefore becomes:

\begin{algorithm}[H]
    \caption{Modified MAPLE Algorithm for Transient Problem Ranking Integration}
    \begin{algorithmic}
    \State \textbf{Data:} Student problem correctness $C_{a,k}$, normalization factor $\alpha$, question $q_a$
    
    \State Get student correctness $C_{a,k}$ after solving question $q_a$
    \If{$C_{a,k}$ is $1$}
        \State Increase weights for questions more difficult than $q_a$:
        \State $m_a = \alpha m_a$
        \State Decrease weights for questions less difficult than $q_a$:
        \State $m_a = \frac{1}{\alpha} m_a$
    \Else
        \State Decrease weights for questions more difficult than $q_a$:
        \State $m_a = \frac{1}{\alpha} m_a$
        \State Increase weights for questions less difficult than $q_a$:
        \State $m_a = \alpha m_a$
    \EndIf
    \end{algorithmic}
\end{algorithm}

\section{Results: Student Simulation} \label{sec:simulations}
To validate our hierarchical MAB architecture before real-world implementation, we utilized BKT to simulate a roster of 1500 students in an adaptive learning environment (\citeauthor{andersonBKT}, \citeyear{andersonBKT}; \citeauthor{badrinathPyBKT}, \citeyear{badrinathPyBKT}; Appendix \ref{pybkt-explaination}). First, we fitted our hierarchical MAB application for our organization's AI material, where five of our education sections were utilized for the MAB intelligent tutoring. These five sections were specifically picked as they have multiple concepts incorporated into each section, versus the other shorter sections with only one concept each, which ensured the algorithmic results captured are indeed hierarchical.

As the data available from our organization's course material was not extensive enough for accurate result interpretation, we chose to transform the ASSISTments dataset (\citeauthor{wangAssistments}, \citeyear{wangAssistments}) by mapping their data to our own concepts and questions to train the BKT model with. Question difficulty was deciphered by calculating the inaccuracy rate to each problem and adapting it linearly ($d_a= 4 \cdot \text{inaccuracy rate}_a + 1$) to obtain a difficulty score within our $[1,5]$ scale (Section \ref{difficulty_score}).

We defined three groups for simulation: one with a randomized question sequence (where a question is picked at random from a section's question bank), one with a difficulty-agnostic hierarchical MAB sequence (where problem difficulties are not utilized), and one fully realized hierarchical MAB sequence with problem difficulties included.  Each group contained 500 simulated students. We simulated the hierarchical MAB students to the algorithms’ completions and ran the randomized question sequence algorithm for the exact number of questions that was previously required by the difficulty-agnostic hierarchical MAB sequence.

\begin{figure}
\begin{center}
\includegraphics[width=0.90\linewidth]{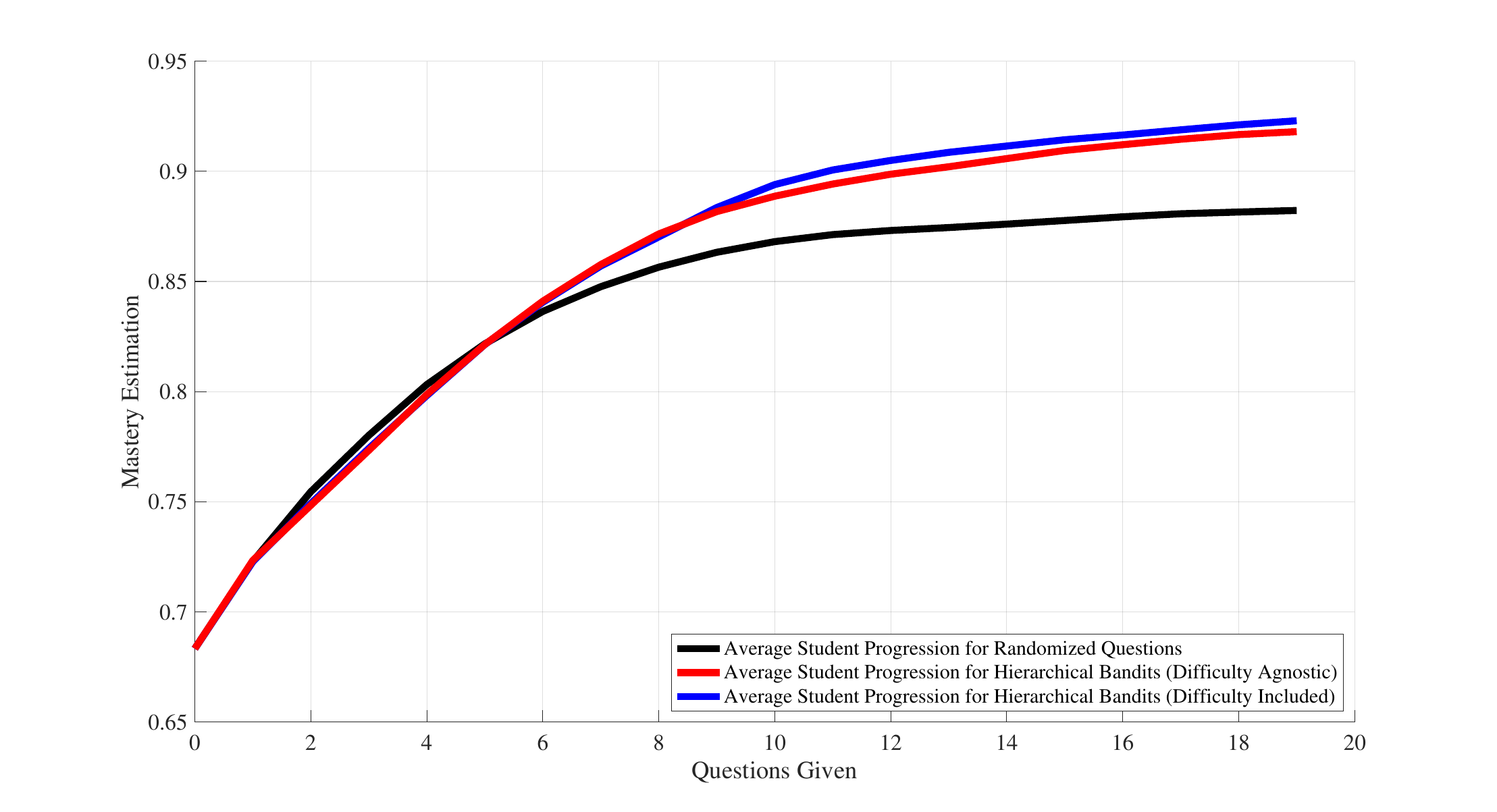}
\caption{\label{fig:mabBKTsim}The average progression for groups of 500 simulated students' mastery of 5 sections (15 concepts) of material, using BKT. This includes a randomized sequence of questions (black), our hierarchical multi-armed bandit framework without problem difficulty considerations (red), and our hierarchical multi-armed bandit framework with problem difficulty included (blue).}
\end{center}
\end{figure}

Figure \ref{fig:mabBKTsim} depicts each of the $500$ simulated student groups' averaged progression (in their groups of 500) resultant from each algorithmic sequence recommendation. On average, we observe that, initially, the randomized sequence performed best from our mastery estimation. This is likely attributed to the inherent advantage of randomization, which boosts broad coverage of concepts rather than revisiting previously addressed ones. As such, this approach likely increased the probability of attaining more correct answers from different concepts, thereby contributing initial, significant exponential gains to the mastery estimation\footnote{Problems typically have response correctness probabilities greater than $50\%$. This, in turn, promotes problem coverage yielding greater mastery estimation initially as mastery estimation contributions towards individual concepts exponentially decays with successive correct answers.}. 

However, after a certain problem-prompting threshold, the hierarchical MAB frameworks suggest more reliability in bringing students above the randomized sequence’s average asymptotic mastery level. This likely occurs as the MAB algorithms can better detect students struggling with certain concepts and therefore can give more problem suggestions within that concept. We witness that students obtained a slightly higher mastery level with the difficulty-included hierarchical MAB sequence as opposed to the difficulty-agnostic one, demonstrating the isolated effect of difficulty incorporation. 

\section{Conclusion and Future Work}
In this paper, we present a deployable, open-source, and state-of-the-art MAB intelligent tutor for remote education. We combine prior research into MAB intelligent tutoring literature to synthesize our algorithm, which creates pedagogical advantages for instructors like assimilating concept maps and problem difficulties. Our algorithm utilizes hierarchical MABs which contain separate MAB agents to select concepts and problems. Lastly, we perform BKT simulations to build evidence for our algorithm’s efficacy in real-world educational environments, which contrasts a randomized, MAB difficulty-agnostic, and MAB difficulty inclusion problem sequencing.

Future work includes the need for real student trials that aren’t dependent on student knowledge models or idealized conditions. This invites the possibilities of other considerations, such as real-time self-updating of problem difficulties, window size trade-off investigation, and material redirects for underperforming students, to name a few.

\bibliographystyle{rlc}
\bibliography{main} 

\section{Appendix}

\subsection{ZPDES, MCM, and MAPLE Parameter Choices}
\label{zpdes-maple-hypertune}

We obtained many of our ZPDES and modified MAPLE algorithm initial parameters and hyperparameters through tuning before student simulation to converge onto values that a) progress students at a reasonable rate and b) avoid overvaluing nor neglecting both complex and simple problems. For ZPDES, we obtain $\gamma=0.1$, $w_{a,0}=0.5$, $\xi=7.37$, $L_{concept} = 4$, $L_{problem} = 2$, $h =  0.74$. Problems unlocked by completing prerequisites are also appended weights of $w_{a,1}=2$. For MAPLE, we obtain $\alpha = 1.3$.

For the MCM algorithm, \cite{pashlerMCM} used mass simulations for optimizing $\xi_i$ and $\tau_i$. However, our lack of prior question data for model fitting leads us to follow \cite{muMABtheory}'s implementation (who had an identical dilemma for parameter value choice) of choosing $\xi_i = 1$ and $\tau_i = i$. Given the brevity of our platform's sections (Appendix \ref{pybkt-explaination}), we neglect tuning a memory threshold $m_t$ or a memory multiplier $m_m$ value for our individual application.

\subsection{Miscellaneous MAB Modifications}
We modify the algorithm so that the belief states for problems successfully completed by the student are marked as mastered and therefore are not given to the student again. Furthermore, if all problems in a problem MAB instance have been completed, the conceptual MAB will have the corresponding concept's belief state updated to mastered as no more problems within the given concept are available to give to the student. \label{one-time-correctness}

\subsection{MAPLE Modifications List}
\label{sec:maple-changes-made}

The following modifications were made to the Update Question Grade calculation step of the MAPLE algorithm for integration with our hierarchical MAB algorithm:

\begin{itemize}\itemsep0em 
    \item Instead of updating the weight problem weight $w_a$ directly, we choose to update the multipliers for each problem $m_{a}$.
    \item Our learning environment assumes that only binary grades are possible ($0$ and $1$) for the student grade $g_s$. Therefore, we lower the number of hyperparameters by removing the passing grade threshold conditional $g_s > \eta$ and replace it with $C_{a,k} == 1$, that is, if the exercise given at time $k$ is answered correctly.
    \item Only one normalization factor $\alpha$ is used and instead the normalization factor used for lowering multipliers ($\alpha_3$ in \citeauthor{segalDifficultyMAB}, \citeyear{segalDifficultyMAB}) is the inverse of $\alpha$.
    \item If the last question was answered correctly, we also decrease the multipliers for questions less difficult than $q_a$ by the inverse of $\alpha$. 
    \item If the last question was answered incorrectly, we also increase the multipliers for questions less difficult than $q_a$.
    \item We remove direct exploration rate $\gamma$ calculations from MAPLE.
    \item As each problem can only be answered correctly one time (see Appendix \ref{one-time-correctness}), we choose to remove the exponential reward multiplier ($e^R$ in \citeauthor{segalDifficultyMAB}, \citeyear{segalDifficultyMAB}) from the algorithm.
\end{itemize}
By changing the problem multipliers for problems easier than the last given question, the algorithm counteracts the initial problem difficulty integration incurred by Equation \ref{eq:initialProblemDifficultyIntegration}. However, as all problems (other than those of equal difficulty to $q_a$) now have weights updated according to difficulty, it is no longer necessary to change the exploration rate $\gamma$.

\subsection{Hierarchical MAB Schematic}

\label{sec:mab_schematic}
\begin{figure}[H]
\begin{center}
\includegraphics[width=0.90\linewidth]{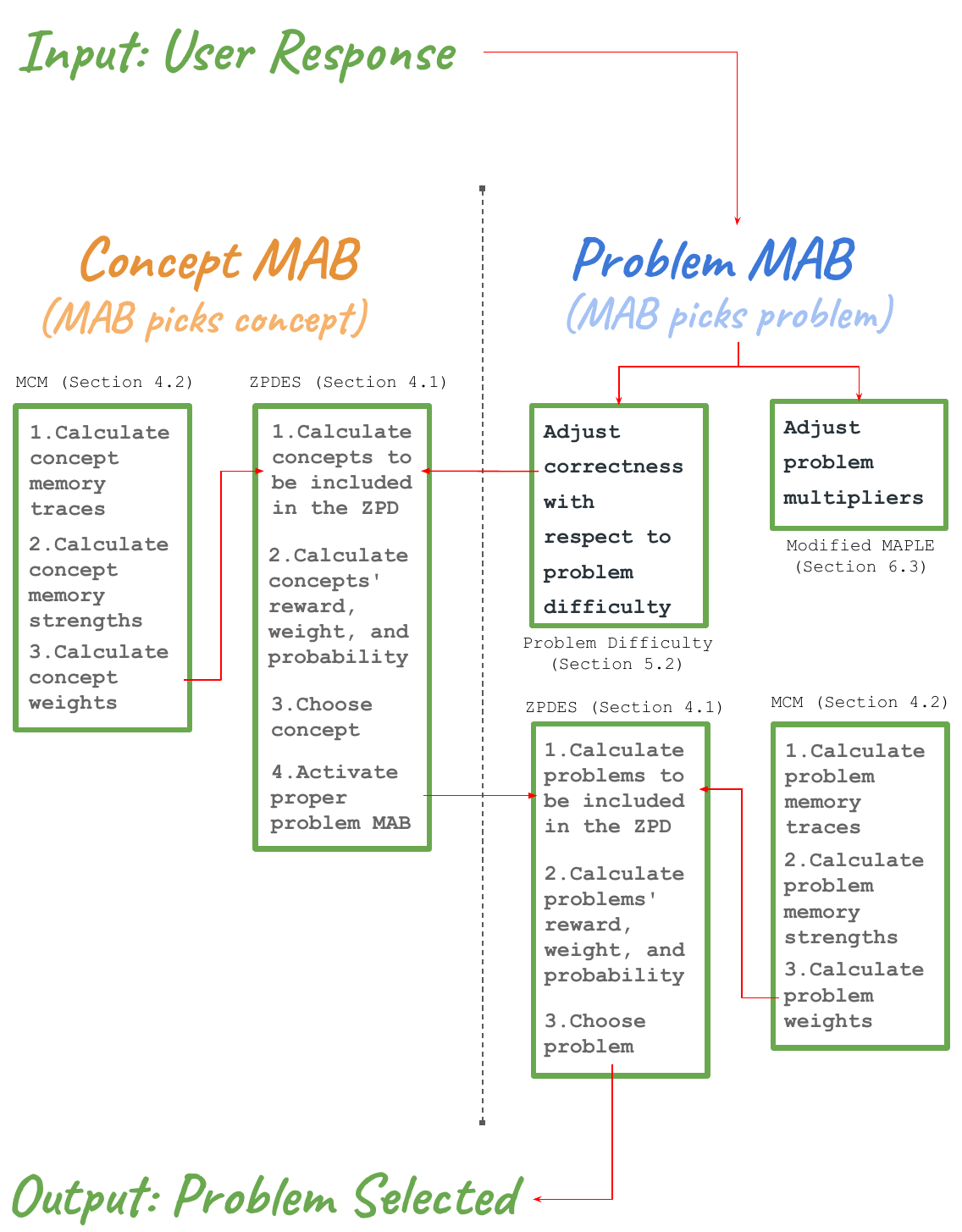}
\caption{\label{fig:mab_schematic}A schematic of how the concept MAB and problem MABs interact.}
\end{center}
\end{figure}

\subsection{pyBKT Setup}
\label{pybkt-explaination}

The pyBKT model implements a Hidden Markov Model (HMM) with sequences of the students' response correctness as observable nodes and students' latent knowledge throughout proceeding knowledge states as hidden nodes. The model trains on past students' response history to fit the HMM's learn (probability of transmission from an unlearned state $\lambda_t=0$ to a learned state $\lambda_{t+1}=1$), prior (probability of initial learned state $\lambda_0=1$), guess (probability of responding correctly despite a presently unlearned state $\lambda_t=0$), and slip (probability of responding incorrectly despite presently learned state $\lambda_t=1$) parameters (\citeauthor{badrinathPyBKT}, \citeyear{badrinathPyBKT}).

When an unlearned latent state is present ($\lambda_t=0$), we provide a binary correctness $C_{t,k}=1$ at the current guess probability and otherwise respond $C_{t,k}=0$. Conversely, when a learned latent state is present ($\lambda_t=1$), we provide correctness $C_{t,k}=0$ at the current slip probability and otherwise respond $C_{t,k}=1$ (identical to \citeauthor{muMABtheory}, \citeyear{muMABtheory}). 

We do not implement any form of forgetting (learning states can not degrade from $\lambda_t=1$ to $\lambda_{t+1}=0$) in our simulations to simplify our assumptions, doing so in part to our observed low question attempt count for section completeness (see Figure \ref{fig:mabBKTsim}). Consequently, we neglect the MCM algorithm's weight contributions for this experimentation.

\end{document}